# An Earthquake Can Be Predicted

Manana Kachakhidze[*1], Nino Kachakhidze-Murphy[1]

[1]Georgian Technical University, Natural Hazard Scientific-Research Center
0171, Georgia, Tbilisi, Kostava str. 77
[*]Corresponding author: Manana Kachakhidze kachakhidzem@gmail.com

**Abstract**

The present consolidated paper represents the VLF/LF electromagnetic radiation as the earthquake's true precursor. The search mainly is carried out on the basis of a theoretical model of the generation of electromagnetic emissions during the earthquake preparation period and earthquake prediction methodology. It is shown that this parameter is capable of describing the fault formative process in the focal area. Besides, VLF/LF electromagnetic radiation frequency analysis gives the possibility simultaneously to determine all three characteristic parameters necessary for incoming earthquake prediction (magnitude, epicenter, and time of occurring).

It is shown that the prediction of moderate and strong earthquakes is possible with great precision.

**Introduction**

Earthquake-stricken humanity constantly has a question: cannot earthquakes be predicted?

Professor Max Wyss has answered this question with hope: I doubt that human curiosity and ingenuity can be prevented in the long run from exploring fully the extent to which at least some earthquakes are predictable, although it is not easy (20).

We agree with Professor Max Wyss that earthquake predictability is not easy, but we already emphasize that earthquakes are predictable.

In this article, we want to briefly introduce the readers to the novelties achieved in the last decades in the field of earthquake research, which leads to earthquake prediction.

Studies that appeared in the scientific literature since the end of the last century confirm that geophysical phenomena, which may accompany the earthquake preparation process and expose themselves several months, weeks, or days prior to earthquakes, take place in the seismogenic area. They are VLF/LF electromagnetic emissions, changing of the intensity of the electro telluric current in the focal area, perturbations of the geomagnetic field in forms of irregular pulsations or regular short-period pulsations, perturbations of the atmospheric electric field, irregular changing of characteristic parameters of the lower ionosphere (plasma frequency, the height of D layer, etc.), irregular perturbations reaching the upper ionosphere (namely F2-layer, for 2-3 days before the earthquake), increased intensity of electromagnetic emissions in the upper ionosphere in several hours or tenths of minutes before the earthquake, lighting before the earthquake, infrared radiation, total electron content (TEC) anomalies, and etc. It is obvious that all the above-mentioned phenomena are not observed before every earthquake and do not have a place in the noted sequence.

These facts confirm that these anomalous perturbations are related to the earthquake preparation process, so it is not excluded that any of these fields may be an earthquake true precursor.

An earthquake is a geological phenomenon. A well-known avalanche-unstable geological model of fault formation describes (15) the origination of different size cracks and finally, the main fault length formation process in the focal area of an incoming earthquake. The sizes of cracks available for failure in earthquakes are fractally distributed (20).

If the geophysical field is a true precursor, it must accurately describe the geological model mentioned above, and in this regard, must analytically explain the complex process of fault formation from the beginning of the appearance of microcracks to the final formation of the main fault and the restoration of equilibrium in the medium, not only qualitatively, but quantitatively too.

Should be noted that discussing the earthquake problem, the scientist has no right to solve the task by allowing certain boundary conditions because it will not give us a real result, especially the possibility of predicting an earthquake.

Let's say in advance that we were able to find the field being an earthquake's true precursor. It is the VLF/LF electromagnetic emissions detected prior to earthquakes.

This opportunity has been given to us by the research results conducted by the world-famous scientists, whose works have been going on since the end of the last century up today. On the one hand, according to these works:

- 1) EM emissions appear approximately several weeks before the earthquake. 2) The spectrum of electromagnetic radiation is characterized by the following sequence: MHz, kHz. 3) These emissions are accompanied by ULF radiation. 4) VLF/LF electromagnetic emissions before the earthquake become very weak or completely disappear (so-called "silence" appears). 5) The "silence" of VLF/LF EM radiation is followed by an earthquake (1, 2,3, 6,7,8,9,10,16,17).

- On the other hand, for detecting EM radiation that existed prior to the earthquake, the relevant radio networks were established by Japanese researchers in the 2000s (Japanese-Pacific VLF/LF Network) and a European network (INFREP) (3) by Japanese, Russian, and Italian teams' cooperation. It is worth underling that there are examples of strong earthquakes when no radiation has been detected (4).

In 2010 an article was published (16) where the evolution of EM emissions under real conditions have been shown in the case of the L'Aquila earthquake.

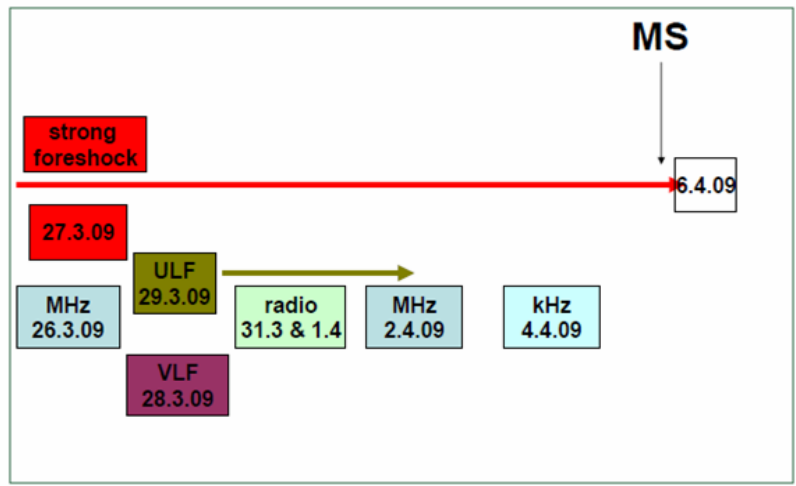

Fig. 1. Evolution of EM emissions in case of L'Aquila earthquake

This picture combines the studies pointed out above during the current process in the pre-earthquake period (27.03.2009-6.04.2009). It confirms the existence of EM radiation before an earthquake and its frequency changing with some regularity (1,2,3,6,7,8,9,10,16,17).

However, the possibility to predict an earthquake was not seen.

Hence the necessity for studying the immediate cause of the generation of EM emissions detected prior to earthquakes became the primary issue for us.

**Discussion**

Since the sizes of cracks are fractally distributed, during the studies, we had to take into account that the seismogenic area is an oscillatory-distributed system.

A seismogenic zone can be considered to be distributed system because the mass, elasticity (of the mechanical system), capacity, and inductance (of the electric system) are roughly uniformly spread throughout the entire in the whole volume of the system.

Obviously, the mass and elasticity characteristic of the mechanical system changes in the earthquake preparation area, but the corresponding mechanical field reflecting these processes on the Earth's surface is not observed during the earthquake preparation process.

Therefore, we must consider the earthquake preparation zone as an electrically distributed system, where each small element has its own capacitance and inductance. In addition, as mentioned above, the changes in the frequency of electromagnetic radiation in the process of earthquake preparation are characterized by a certain regularity.

If it turns out that these changes have been caused by changes in each of the small element's own capacity and inductance (since these changes mainly take place in the fractals during the main fault formation), then obviously electromagnetic radiation could be considered a true precursor.

As a result of relevant searching it was created the model (12).

In the referred work it is obtained the formula (1), which analytically connects with each other the main frequency of the observed electromagnetic emissions and the linear dimension of the emitted body:

$$\omega = \beta \frac{c}{l} \qquad (1)$$

where β is the characteristic coefficient of geological medium (it approximately equals 1).

The question is what is the emitted body.

As known, the principal process taking place in the focal area during the earthquake preparation period is the main fault formative process which is accompanied by EM emissions (1,2,3,6,7,8,9,10,16,17) By the model (12), the same medium during the earthquake preparation period is considered an oscillation system, and it is obtained the formula (1) of the self-generated frequency. Therefore, it is clear that the emitted body is the main fault and its length is $l$.

According to the formula (1) of determination of fault length by the observed frequency, it is possible in advance, before the earthquake, to calculate one of the important characteristic of an incoming earthquake, magnitude, by formulas of dependence of magnitude on the fault length.

For example, in the case of the L'Aquila earthquake, the frequency of electromagnetic radiation equal to about 20 kHz was fixed (18). According to the formula (1) $l \approx 15$ km. By the formula (2) (19)

$$lg\, l = 0.6 Ms - 2.5 \qquad (2)$$
$$M = 6.1$$

If we compare the obtained result with the real data (5), where $l \approx 16$ we can say for sure that according to the formula (1), obtained $l$ is the length of the backbone fault of the earthquake and, therefore, the emitted body really is the main fault.

This means that a changing of the geophysical field characteristic parameter (VLF/LF EM emissions frequency) is inevitably related to changes in the fault length in the focus, which allows us not only qualitatively but also quantitatively to evaluate the complete picture of an earthquake preparation from the appearance of microcracks in the focal area, up to the formation of the main fault and to a final equilibrium state. Hence the pre-earthquake VLF / LF electromagnetic radiation turned out to be a true precursor.

On the other hand, let's remind of the classical geological model - an avalanche-like unstable model of fault formation in the focal area.

It is logical to assume that: If VLF / LF electromagnetic radiation is a true precursor, it must be an exact manifestation of the mechanical stages of the avalanche-unstable geological model.

Let's check this issue as well:

It is known that the avalanche-like unstable model of fault formation is divided into four main stages: In the first stage, which can go on for several months throughout the whole seismogenic area, the chaotic formation of micro-cracks without any orientation takes place (15). This stage of formation of micro-cracks is the reversible process - at this stage not only micro-cracks can be formed but also the so-called "locked" can occur. Cracks created at this stage are small. This stage, according to our model (formula 1) in the electromagnetic emissions frequency range, should be expressed by the discontinuous spectrum of electromagnetic radiation in MHz diapason (12), which is proved by the latest special scientific works (6,7,10,16).

The second stage of the avalanche-like unstable model of fault formation is an irreversible avalanche process of already somewhat oriented microstructures, which is accompanied by the inclusion of the earlier "locked" sections.

Based on our model (12), we have to suppose that this stage in the emissions frequency spectrum should be already expressed by MHz continuous spectrum. Although, because the lengths of the cracks start to increase at the expense of aggregation of primary small cracks, the values of electromagnetic emissions frequency must gradually decrease.

According to the avalanche-like unstable model, this process takes place a few days before the main shock (7,16). The transition of the MHz emissions in kHz in the frequency spectrum of electromagnetic radiation, according to formula (1), corresponds to the very stage when the crack length already reaches about a kilometer (7,12,,13,16,18).

In the third stage of the avalanche-like unstable model of fault formation, the relatively big size faults unite into one - the main fault. This process should correspond to the gradual fall of frequencies in kHz (12), which according to formula (1) refers to the increase of fault length in the focal area. By (2) formula, it also refers to the increase of magnitudes of the expected earthquake (19).

Of course, an association of cracks into one fault, which at the final stage of earthquake preparation proceeds intensely, will use a definite part of energy accumulated in the focal area and therefore, will result in its decrease (11).

In such a situation, a period settles before an earthquake (which can last from several hours to even some days), when a fault is already formed, while an earthquake has not occurred yet, since accumulated tectonic stress is not yet sufficient to overcome the limitation of the solidity of the geological environment.

The system, which is waiting for a further "portion" of tectonic stress, is in the so-called "stupor - waiting" condition, in the principle, the process of main fault formation is not going on in it anymore, and respectively, electromagnetic emissions would not take place. This is proved by experiments too (9,10,16).

This process is expressed correspondingly in the electromagnetic emissions spectrum: some hours before the earthquake (up to 2 days) in the spectrum the emissions interruption or "electromagnetic emissions silence" is observed (16). Of course, we should expect the renewal of electromagnetic emissions exactly before the earthquake (the fourth stage of the fault formation process).

In the period of electromagnetic emissions monitoring the moment of emissions spectrum interruption is urgent for determining of time of occurrence of an incoming earthquake (the second characteristic parameter of the incoming earthquake), since at the final stage of earthquake preparation, as it is mentioned above, a very short time is needed to fill in the critical reserve of tectonic stress for main fault realization.

Thus, good conformity and synthesis of our model and avalanche-like unstable model of fault formation are evident and rather harmonious.

VLF/LF EM radiation is useful for the determination of the epicenter (the third characteristic parameter of the incoming earthquake). From the points selected around the receiver, where VLF/LF EM emissions are fixed, by the Direction-finding method, it is possible to define the incoming earthquake epicenter.

Today, in the scientific literature dealing with the study of the problem of earthquakes, in relation to one or another geophysical field that changes anomalously during the period of earthquake preparation, the term "precursor" is used. Corresponding theories have also been created, but the possibilities of predicting earthquakes are not yet seen.

The question naturally arises: what advantage does VLF/LF electromagnetic radiation have in regards to earthquake prediction?

In the results of studying, we may underline the advantage of VLF/LF electromagnetic radiation: VLF/LF EM emissions turned out to be the unique precursor because it gives the promising possibility of simultaneous determination of moderate and strong, inland incoming earthquake magnitude, epicenter, and time of occurrence. The revealed precursor is the first and only precursor, which describes the fault formation process in the incoming earthquake focus. Besides, it numerically calculates fault length (magnitude) at any moment of monitoring.

It is worth underlying that no reliable criterion is in seismology yet, can distinguish the strong foreshocks from the mainshock. This issue can be resolved with rather a high accuracy on the basis of a theoretical model (12): if after any shock electromagnetic emissions still continue to exist and the frequency data still tend to decrease, it means that the process of fault formation in the earthquake focus is not completed yet and we have to wait for the mainshock. It is on the contrary in the case of aftershocks. This proves once again that VLF/LF electromagnetic radiation is the true precursor.

In the next step, our interest was to "see" main fault formation and EM radiation so-called "silence" signs in EM radiation retrospective records of a real earthquake. Besides, we wanted to try to determine the magnitude i.e. our aim was to check our theoretical searching with real continuous data. In other words, we would like to "predict" an already occurred earthquake.

In order to accomplish this task, we worked out retrospective data of INFREP (European Network of Electromagnetic Radiation) for the Crete earthquake with M= 5.6 (25/05 / 2016, 08:36:13 UTC). Studies have been conducted with continuous data for 73 days (Kachakhidze, et.al, 2019) (14).

INFREP network fixes every minute amplitudes of 10 different baseline frequencies of VLF/LF electromagnetic radiation in diapason 20 270 Hz - 270 000 Hz.

It is clear that, if any frequency channel actually reflects the earthquake preparation, the avalanche-like unstable geological process should be reflected in the frequency data (formula 1).

For this reason, we calculated the lengths of cracks in every minute corresponding to all frequency channels towards the lengths relevant to the channel's baseline frequencies (in the percentage). It was found out two active channels (37 500 Hz) and (45 900 Hz) as the average daily value of the cracks lengths was the maximum for them. This means that from discussed 10 channels, only two, with the above-mentioned frequencies, described the earthquake preparation process. In this case, according to the above given formula (1), the magnitude of the incoming earthquake should be between 5.5 and 5.7 (Crete earthquake magnitude is really estimated as M= 5.6).

Besides, we found out that date diurnal periodic variations of electromagnetic emissions are clearly expressed on all channels, except the (37 500 Hz) channel. We have such variations in the (37 500 Hz) channel recordings too, but till some period, namely up to 02.05.2016, after which the anomalous process starts, indicating that the avalanche–unstable process of fault formation has already begun (Fig. 2).

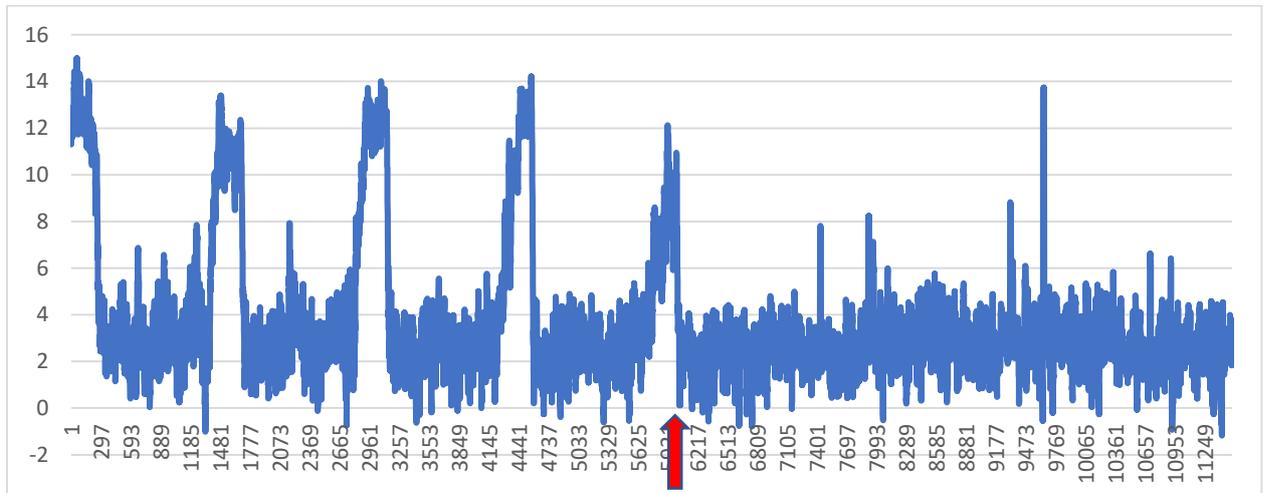

Fig.2. The initial moment of the avalanche-like unstable process of fault formation (the red arrow)

Since, in the case of discussed earthquake, only (37 500 Hz) frequency channel meets both conditions: for this channel the average daily value of the lengths of the cracks is maximal and an avalanche process of fault formation appears only on it, obviously, to predict the earthquake, we must rely only on the data of this channel.

During analyses, has been found, that 19 days before the earthquake, the fault formation avalanche process appeared in the (37 500 Hz) frequency channel. By the formulas (1, 2), the expected length of the main fault is about 8 000 meters, and the magnitude approximately is equal to 5.6.

In order to analysis of the full process of earthquake preparation, we elaborated the daily averaged frequencies by using the average square deviation method, and we calculated $\bar{x} \pm \sigma$ values (fig.3).

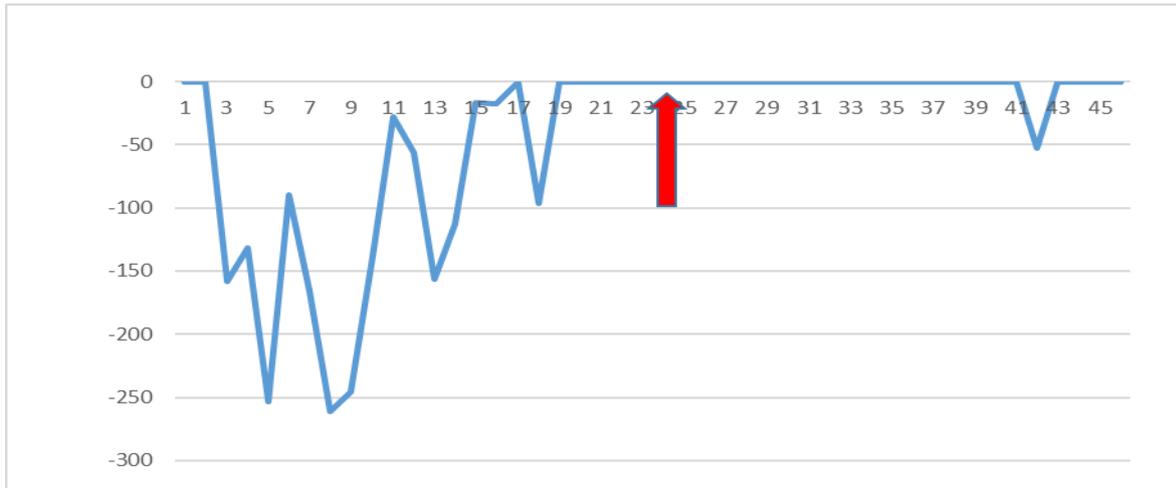

Fig. 3. The avalanche process of fault formation and EM emissions "silence"
period before the considered earthquake (with the red arrow is shown earthquake occurring moment)

It is important that such elaboration of the data manifests the moment of the onset of EM emission "silence" prior to an earthquake. it was equal to about two days in the case of the Crete earthquake.

Thus, a real earthquake's retrospective data analysis shows the possibilities to fix the exact moment of the beginning of the fault formation avalanche process and EM radiation's "silence" onset moment. In other words, in the case of the EM emissions monitoring, we have the possibility to observe the whole process of the earthquake preparation and finally, predict the earthquake.

It is clear that, when the tectonic stress increases in a certain zone of the region, physical, chemical, electromagnetic, mechanical, etc. fields will change in this area, most of which will be observed on the earth's surface in the form of abnormal changes. The features of these fields in connection with earthquakes are revealed in the fact that they usually, only qualitatively express the processes in the focal area. Based on them it is not a possible simultaneous determination of all three parameters necessary for incoming earthquake prediction (magnitude, epicenter, and time of occurring). It means, that these fields really have only indicative features in the process of earthquake preparation.

It is obvious also, that the process of fault formation without fail must be reflected in the abnormal changes in the certain geophysical field and this phenomenon should be recorded prior to the earthquake.

This means that a change in the characteristic parameter of this certain geophysical field will inevitably be related to changes in the length of the fault, which will allow us not only qualitatively but also quantitatively to evaluate the complete picture of an earthquake preparation process from the appearance of microcracks in focus, up to the formation of the main fault and to a final equilibrium state.

The prediction of the incoming earthquake will be possible only by analyzing an anomalous change in such a field.

Finally, based on current scientific studies, anomalous changes in the fields, observed in the seismogenic area during the earthquake formative period, must be divided into three groups: earthquake triggering factors, earthquake indicators, and earthquake precursors. These three factors can be defined as follows:

1. The physical field that exists independently of the earthquake preparation process should be considered as an earthquake trigger (natural or man-made) if at the final stage of earthquake preparation when tectonic stress reaches the limit of the solidity of the geological medium, it can affect the tectonic stress and correct the time of earthquake occurrence (for example, tides, in some cases changing of atmospheric pressure, reciprocal - influences of earthquakes, sharp changes of water level in reservoirs, etc.).

2. An indicator of earthquakes can be called a physical field, that exists independently of the earthquake formative process, but due to the existence of this process, it undergoes abnormal changes from a certain period of earthquake preparation. It should be emphasized that the origin of the earthquake indicator is not caused by the direct process of formation of the main fault length. The indicator is caused by perturbations of geophysical fields in seismogenic (mostly focal) areas (e.g. earth magnetic field anomalies, changes of atmospheric electric field potential gradient, telluric field anomalies, gases emissions from rocks, TEC anomalies and etc.)

Mainly, in the earth-atmosphere-ionosphere coupling system, in the case of perturbations, which may be caused not only by terrestrial but also by solar reasons, the same geophysical field often manifests "indicator" properties in case of the presence of both of these factors (e.g. telluric field anomalies are observed both in the process of earthquake preparation and in bad weather, TEC anomalies reveal themselves in both strong earthquakes and thunderstorms, the earth's magnetic field anomalies are fixed in both earthquakes and magnetic storms, changings in the electric potential of the earth's surface appear as in the case of bad weather but earthquakes as well, etc.).

Therefore, "earthquake indicators" can be called those anomalous changes of geophysical fields, which are really caused by the process of earthquake preparation, but they cannot analytically describe the process of main fault formation, that is, these fields are not useful to predict the earthquake.

3. The earthquake true precursor is the physical field, the cause of which origination is the process of the formation of the main fault length in the medium and the anomalous changing of which parameters (or parameter) expresses the process of formation of the main fault final length at the expense of the arising and coalescence of the cracks due to the accumulation of tectonic stress.

That is, by means of precursor, it will be possible quantitatively to monitor changes in the length of the main fault from the beginning of the fault formation to the complete ending of the earthquake process.

We would like to emphasize that the true precursor, VLF/ LF EM radiation, fully meets the Guidelines for Submission of Earthquake Precursor Candidates (21).

In case of weak earthquakes, we have to wait for the electromagnetic emissions in high frequency diapason but high-frequency waves attenuate rapidly, and observing them on the earth's surface is difficult.

**Conclusion:**

In the results of the research the following conclusions are made:

1) By the active channel frequency, it is possible to determine the length of the "cracked backbone" on which the process of cracks origination is going on actively and ultimately the main fault forms.

2) By the length of the "cracked backbone" it is possible to determine the magnitude of an incoming earthquake with certain accuracy about several dozen days before the earthquake.

3) After the active frequency channel detection, it is already possible to determine the incoming earthquake epicenter with certain accuracy.

4) In order to the short-term prediction of a moderate and strong earthquake, it is recommended to begin careful monitoring of the frequency data from the beginning of the fault formation avalanche-unstable process, keep an eye on the process dynamics, and fix the starting moment of "silence" of the electromagnetic radiation.

In the case of monitoring electromagnetic radiation, it is possible to make a prediction of an incoming earthquake from several dozen up to several hours before earthquake occurrence (depending on the geological medium).

5) It is possible to separate the foreshock (aftershocks too) from the main shock.

Thus, VLF/LF EM emissions turned out to be the true precursor, which gives the possibility of simultaneous determination of moderate and strong, inland incoming earthquake epicenter, time of occurrence, and magnitude with high precision.

It is worth underlining that the revealed precursor is the first and only among other precursors, which describes the fault formation process in the incoming earthquake focal area and numerically calculates fault length (magnitude) at any moment of monitoring.

Thus, to the question "is it possible to predict earthquakes?" we may answer that moderate and strong earthquakes can be predicted.

**Acknowledgments:** The authors are grateful to the network INFREP, for providing us with the MHz and kHz fractal-electromagnetic emissions data used in this paper.